\documentclass[global]{svjour}
\usepackage{amssymb}
\usepackage{amsfonts}
\usepackage{amsmath}

\input{tcilatex}

\begin{document}

\journalname{MPLA}
\title{Quasinormal modes of charged dilaton black holes and their entropy
spectra}
\author{I. Sakalli\inst{1}}
\institute{Department of Physics, Eastern Mediterranean University, Gazimagosa, North
Cyprus, Mersin 10, Turkey. 
\email{izzet.sakalli@emu.edu.tr}
}
\dedication{}
\offprints{}
\mail{}
\maketitle

\begin{abstract}
In this study, we employ the scalar perturbations of the charged dilaton
black hole (CDBH) found by Chan, Horne and Mann (CHM), and described with an
action which emerges in the low-energy limit of the string theory. A CDBH is
neither asymptotically flat (AF) nor non-asymptotically flat (NAF)
spacetime. Depending on the value of its dilaton parameter $a$, it has both
Schwarzschild and linear dilaton black hole (LDBH) limits. We compute the
complex frequencies of the quasinormal modes (QNMs) of the CDBH by
considering small perturbations around its horizon. By using the highly
damped QNMs in the process prescribed by Maggiore, we obtain the quantum
entropy and area spectra of these BHs. Although the QNM frequencies are
tuned by $a$, we show that the quantum spectra do not depend on $a,$ and
they are equally spaced. On the other hand, the obtained value of
undetermined dimensionless constant $\epsilon $ is the double of
Bekenstein's result. The possible reason of this discrepancy is also
discussed.
\end{abstract}

\keywords{Quasinormal Modes, Entropy Spectrum, Charged Dilaton Black Holes,
Zerilli Equation, Hurwitz-Lerch Zeta Function, Confluent Hypergeometric
Function. }

\section{Introduction}

There has already been benefits in studying thermodynamics of black holes
(BHs). This subject is believed to be threshold of the unification of
quantum physics with general relativity, which is the so-called quantum
gravity theory (QGT). The reader may see Ref. \cite{Rovelli} and references
therein for a general review of QGT. However, this theory is still under
construction. Recent decades proved that our intricate universe is far from
being easily understandable. In this regard, QGT is perceived as a master
key which resolves many unanswered questions about the universe. For this
reason, the uncompleted form of QGT always stimulates the theoretical
physicists for studying on it more and more.

The starting point of QGT dates back to seventies in which Bekenstein
proposed that BH entropy is proportional to area of \ BH horizon and the
area is quantized \cite{Bek1,Bek2}. Then Bekenstein \cite{Bek3,Bek4,Bek5}
also proved that the BH horizon area is an adiabatic invariant, and
according to Ehrenfest's principle it has a discrete and evenly spaced
spectrum

\begin{equation}
\mathcal{A}_{n}=\epsilon n\hbar =\epsilon nl_{p}^{2},\text{ \ \ \ \ \ \ }%
(n=0,1,2.......),  \label{1}
\end{equation}

where $\mathcal{A}_{n}$\ denotes the area spectrum of the BH horizon and $n$
is the quantum number. Therefore, the minimum increase of the horizon area
is $\Delta \mathcal{A}_{\min }=\epsilon \hbar $ which can be obtained by
absorbing a test particle into the BH. In units with $c=G=1,$ the
undetermined dimensionless constant $\epsilon $ is considered as the order
of unity. Bekenstein proposed that the BH\ horizon is formed by patches of
equal area $\epsilon \hbar $, and moreover professed that $\epsilon =8\pi $.
Motivated by this proposal, many works have been made in this subject in
order to compute the entropy spectrum of various BHs. Different spectra with
different $\epsilon $ have also been presented (see for instance Ref.~\cite%
{Jia} and references therein). One of the significant contributions in
quantizing the entropy of a BH was done by Hod \cite{Hod1,Hod2} who
suggested that $\epsilon $\ can be determined by using the QNM of a BH. As
it is well-known, this mode is the characteristic sound of a BH. Based on
Bohr's correspondence principle (a reader may refer to Ref. \cite{Bohr}),
Hod conjectured that the real part of the asymptotic QNM frequency ($\omega
_{R}$) of a highly damped BH is related to the quantum transition energy
between two quantum levels of the BH. Thus, this transition frequency gives
rise to a change in the BH mass as $\Delta M=\hbar \omega _{R}$.
Particularly for the Schwarzschild BH, Hod computed the value of the
dimensionless constant as $\epsilon =4\ln 3$. Later on, Kunstatter \cite%
{Kunstatter} used the natural adiabatic invariant $I_{adb}$ for system with
energy $E$ and vibrational frequency $\Delta \omega $ (for a BH, $E$ is
identified with the mass $M$ ) which is given by

\begin{equation}
I_{adb}=\int \frac{dE}{\Delta \omega }.  \label{2}
\end{equation}

At large quantum numbers, the adiabatic invariant is quantized via the
Bohr-Sommerfeld quantization; $I_{adb}\simeq n\hbar $. By using the
Schwarzschild BH, Kunstatter showed that when $\omega _{R}$ is used as the
vibrational frequency, the Hod' result $\epsilon =4\ln 3$ is reproduced. In
2008, Maggiore \cite{Maggiore} proposed another method that the QNM of a
perturbed BH should be considered as a damped harmonic oscillator since the
QNM has an imaginary part. Namely, Maggiore considered the proper physical
frequency of the harmonic oscillator with a damping term in the form of $%
\omega =\left( \omega _{R}^{2}+\omega _{I}^{2}\right) ^{\frac{1}{2}}$, where 
$\omega _{R}$\ and $\omega _{I}$\ are the real and imaginary parts of the
frequency of the QNM, respectively. In the large $n$ limit or for the highly
excited mode, $\omega _{I}\gg \omega _{R}$. Consequently one has to use $%
\omega _{I}$\ rather than $\omega _{R}$\ in the adiabatic quantity. With
this new identification, for the Schwarzschild BH it was found that $%
\epsilon =8\pi ,$ which corresponds to the same area spectrum of
Bekenstein's original result of the Schwarzschild BH \cite{Vagenas,Medved}.
To date, there are numerous studies in the literature in which Maggiore's
method (MM) was employed (some of them can be seen in Refs. \cite%
{Samp1,Samp2,Samp3,Samp4,Samp5,Samp6,Samp7}).

In this paper, using the MM with the adiabatic invariant expression (2) we
investigate the entropy and area spectra for the CDBH \cite{CDBH}. CDBHs are
such spacetimes that by tuning the dilaton field one can converts the NAF
structure of the spacetime (including LDBH \cite{Clement,MSH}) to the AF
one, which corresponds to the Schwarzschild BH. Our main motivation is to
examine how the influence of dilaton field effects the BH\ spectroscopy. For
this purpose, we first calculate the QNMs of the CDBH and subsequently use
them in the MM. The obtained entropy spectrum is equally spaced and
independent of the dilaton field. On the other hand, here we get $\epsilon
=16\pi $ which means that the equi-spacing does not coincide with the
Bekenstein's result.

The paper is structured as follows. In Sec. 2, we briefly present the CDBH
metric and its basic thermodynamical features. Also, we show that how the
massless Klein Gordon equation reduces to the Schr\"{o}dinger-type equation
which is the so-called the Zerilli equation \cite{Chandra} in the CDBH
geometry. Sec. 3 is devoted to the derivation of QNM of the CDBH by
considering the small perturbations around the horizon. In addition to that,
in a particular case of highly damped scalar modes, we perform the MM for
the CDBH in order to compute the entropy and area spectra of it. Finally,
the summary and concluding remarks are given in Sec. 4.

\section{CDBH and the separation of the massless Klein Gordon equation on it}

In this section we will first present the geometry and some thermodynamical
properties of the CDBH. Then, we will get the radial equation for a massless
scalar field in the background of the CDBH. Finally, we represent how the
radial equation can be converted to the Zerilli equation \cite{Chandra}
which is none other than one-dimensional Schr\"{o}dinger wave equation.

The $4D$ Einstein-Maxwell-dilaton (EMD) low-energy action obtained from
string theory is given by

\begin{equation}
S=\int d^{4}x\sqrt{-g}(\Re -2(\nabla \phi )^{2}-e^{-2a\phi }F^{2}),
\label{3}
\end{equation}

where $\phi $ describes the dilaton field which is a scalar field that
couples to Maxwell field, $a$ denotes the dilaton parameter and $\Re $ is
the curvature scalar. $F^{2}=F_{\mu \upsilon }F^{\mu \upsilon }$ in which $%
F_{\mu \upsilon }$ is the Maxwell field associated with a $U(1)$ subgroup of 
$E_{8}\times E_{8}$ or Spin(32)/$Z_{2}$ \cite{GHS}. Without loss of
generality, throughout the paper we shall use $a>0$.

In 1995, CHM obtained the CDBH solution to the above action in their
landmark paper \cite{CDBH}. By this end, they used a non-constant dilaton
field. Their solution is described by the following static and spherically
symmetric metric%
\begin{equation}
ds^{2}=-f(r)dt^{2}+\frac{dr^{2}}{f(r)}+R(r)^{2}d\Omega ^{2},  \label{4}
\end{equation}

where $d\Omega ^{2}$ is the standard metric on $2-$sphere and the metric
functions $f(r)$ and $R(r)$ are given by

\begin{equation}
f(r)=\frac{1}{\gamma ^{2}}r^{\frac{2}{1+a^{2}}}(1-\frac{r_{h}}{r}),
\label{5}
\end{equation}

and

\begin{equation}
R(r)=\gamma r^{\acute{N}},  \label{6}
\end{equation}

Here $r_{h}$ denotes the event horizon of the CDBH. $\gamma $\ and $\acute{N}
$ are arbitrary real constants. $\acute{N}$ is related to $a$ by

\begin{equation}
\acute{N}=\frac{a^{2}}{1+a^{2}},  \label{7}
\end{equation}

Furthermore, the dilaton field satisfies

\begin{equation}
\phi =\phi _{0}+\phi _{1}\ln r,  \label{8}
\end{equation}

where

\begin{equation}
\phi _{0}=-\frac{1}{2a}\ln \left[ \frac{Q^{2}\left( 1+a^{2}\right) }{\gamma
^{2}}\right] \text{ and }\phi _{1}=\frac{\acute{N}}{a},  \label{9}
\end{equation}

where $Q$ refers to the electric charge. In this case, the solution for the
electromagnetic (\textit{em}) field is found as

\begin{equation}
F_{tr}=\frac{Qe^{2a\phi }}{R(r)^{2}},  \label{10}
\end{equation}

We should emphasize that the magnetically charged version of the CDBH can
also be derived. This is possible with simply replacing $a\rightarrow -a$ in
the field equations obtained from the action (3) and to consider the \textit{%
em} field as $F_{\theta \varphi }=Q\sin \theta $ (it goes without saying
that $Q$ would be referred as magnetic charge) \cite{CDBH}.

CDBH is not vacuum solution since the action (3) includes a static dilaton
fluid which possesses a non-zero energy-momentum. In fact, considering such
a particular fluid model makes the CDBHs so interesting that they are
neither AF nor NAF. As shown in Ref. \cite{CDBH}, the mass of the BH can be
computed by following the quasilocal mass definition of Brown and York \cite%
{BrownYork} as

\begin{equation}
r_{h}=\frac{2M}{\acute{N}},  \label{11}
\end{equation}

We remark also that the horizon at $r=r_{h}$ hides the singularity located
at $r=0$. In the extreme case $r_{h}=0,$\ metric (4) still exhibits the
features of the BH. Because the singularity at $r=0$ is null and marginally
trapped such that it prevents the signals to reach the external observers.
Unlike to the other charged BHs, a CDBH has no extremal limits. In other
words, it has no zero charge limit. First of all, the eponyms of the LDBH
are Cl\'{e}ment and Gal'tsov \cite{Clement}. Metric functions (5) and (6)
correspond to the $4D$ LDBH which is the solution to the EMD theory \cite%
{Clement} in the case of $a=1$ ($\acute{N}=\frac{1}{2}$). Later on, it is
shown that in addition to the EMD theory, LDBHs are available in
Einstein-Yang-Mills-dilaton and Einstein-Yang-Mills-Born-Infeld-dilaton
theories \cite{MSH}. The most intriguing feature of these BHs is that while
radiating, they undergo an isothermal process. Namely, their temperature
does not alter with shrinking of the BH horizon or with the mass loss.
Furthermore, LDBHs can perform a fading Hawking radiation in which the
temperature goes zero with its ending mass when the quantum corrected
entropy is taken into account \cite{Sakalli1}. On the other hand, while $%
a\rightarrow \infty $ ($\acute{N}=1$) with $\gamma =1,$ metric (4) reduces
to the Schwarzschild BH, which is AF as it is well-known.

Surface gravity of CDBH is calculated through the following expression

\begin{equation}
\kappa =\left. \frac{f^{\prime }(r)}{2}\right\vert _{r=rh}=\frac{r_{h}^{(%
\frac{2N}{a^{2}}-1)}}{2\gamma ^{2}},  \label{12}
\end{equation}

where a prime "$\prime $" denotes differentiation with respect to $r$.
Subsequently, one can readily obtain the Hawking temperature $T_{H}$ of the
CDBH (in gravitational units of $c=G=1$ and $\hbar =l_{p}^{2}$) as

\begin{eqnarray}
T_{H} &=&\frac{\hbar \kappa }{2\pi },  \notag \\
&=&\frac{\hbar r_{h}^{(\frac{2\acute{N}}{a^{2}}-1)}}{4\pi \gamma ^{2}}=\frac{%
\hbar r_{h}^{(1-2\acute{N})}}{4\pi \gamma ^{2}},  \label{13n}
\end{eqnarray}

From the above expression, we see that while the CDBH losing its $M$ by
virtue of the Hawking radiation, $T_{H}$ increases for $a^{2}>1$, decreases
for $a^{2}<1$ and is constant (independent of $M$) for $a^{2}=1$ (LDBH).
Therefore, as mentioned before the LDBH's radiation is such a particular
process that the energy (mass, $M$) transferring out of the BH typically
occurs at a slow rate that thermal equilibrium is maintained. The
Bekenstein-Hawking entropy is given by

\begin{eqnarray}
S_{BH} &=&\frac{A_{h}}{4\hbar },  \notag \\
&=&\frac{\pi }{\hbar }R(r)^{2}=\frac{\pi }{\hbar }\gamma ^{2}r_{h}^{2\acute{N%
}},  \label{14n}
\end{eqnarray}

which leads to

\begin{equation}
dS_{BH}=4\frac{\pi }{\hbar }\gamma ^{2}r_{h}^{(2\acute{N}-1)}dM,  \label{15}
\end{equation}

With these definitions, the validity of the first law of thermodynamics for
the CDBH can be proven via

\begin{equation}
T_{H}dS_{BH}=dM.  \label{16}
\end{equation}

In order to find the entropy spectrum by using the MM, here we shall \
firstly consider the massless scalar wave equation on the geometry of the
CDBH. The general equation of massless scalar field in a curved spacetime is
written as

\begin{equation}
\square \digamma =0,  \label{17}
\end{equation}

where $\square $ denotes the Laplace-Beltrami operator. Thus, the above
equation is equal to

\begin{equation}
\frac{1}{\sqrt{-g}}\partial _{i}(\sqrt{-g}\partial ^{i}\digamma ),\text{ \ \
\ }i=0...3,  \label{18}
\end{equation}

Using the following ansatz for the scalar field $\digamma $ in Eq. (17)

\begin{equation}
\digamma =\frac{\rho (r)}{r^{\acute{N}}}e^{i\omega t}Y_{L}^{m}(\theta
,\varphi ),\text{ \ \ }Re(\omega )>0,  \label{19}
\end{equation}

in which $Y_{L}^{m}(\theta ,\varphi )$ is the well-known spheroidal
harmonics which admits the eigenvalue $-L(L+1)$ \cite{Du}, one obtains the
following Zerilli equation \cite{Chandra} as

\begin{equation}
\left[ -\frac{d^{2}}{dr^{\ast 2}}+V(r)\right] \rho (r)=\omega ^{2}\rho (r),
\label{20}
\end{equation}

where the effective potential is computed as

\begin{equation}
V(r)=f(r)\left[ \frac{\acute{N}(\acute{N}-1)}{r^{2}}f(r)+\frac{L(L+1)}{%
\gamma ^{2}r_{h}^{2\acute{N}}}+\frac{\acute{N}}{r}f^{\prime }(r)\right] ,
\label{21}
\end{equation}

The tortoise coordinate $r^{\ast }$ is defined as,

\begin{equation}
r^{\ast }=\int \frac{dr}{f(r)},  \label{22}
\end{equation}

which yields

\begin{equation}
r^{\ast }=-\gamma ^{2}\frac{r^{2\acute{N}}}{r_{h}}\Phi (\frac{r}{rh},1,2%
\acute{N}),  \label{23}
\end{equation}

where $\Phi $ denotes the Hurwitz-Lerch Zeta function (see Ref. \cite%
{Srivastava}). This function is defined by

\begin{equation}
\Phi (z,s,b)=\sum\limits_{k=0}^{\infty }\frac{z^{k}}{\left( k+b\right) ^{s}},
\label{24}
\end{equation}

and $\Phi (\frac{r}{rh},1,2\acute{N})$ can be transformed into the
hypergeometric function as

\begin{equation}
\Phi (\frac{r}{rh},1,2\acute{N})=\frac{1}{2\acute{N}}\text{ }_{2}F_{1}(1,2%
\acute{N};1+2\acute{N};\frac{r}{rh}),  \label{25}
\end{equation}

where $_{2}F_{1}$ represents the Gaussian hypergeometric function. Finally,
it follows from Eq. (23) that

\begin{equation}
\lim_{r\rightarrow rh}r^{\ast }=-\infty \text{ \ and }\lim_{r\rightarrow
\infty }r^{\ast }=\infty .  \label{26}
\end{equation}

\section{QNMs and entropy spectrum of CDBH}

In this section, we intend to derive the entropy and area spectra of the
CDBH by using the MM. Gaining inspiration from the studies \cite%
{Appro1,Appro2,Appro3}, here we use an approximation method in order to
define the QNMs. Since the effective potential (8) diverges at the spatial
infinity ($r^{\ast }\rightarrow \infty $) and vanishes at the horizon ($%
r^{\ast }\rightarrow -\infty $), therefore the QNMs are defined to be those
for which we have only ingoing plane wave at the horizon, namely,

\begin{equation}
\left. \rho (r)\right\vert _{QNM}\sim e^{i\omega r^{\ast }}\text{ at }%
r^{\ast }\rightarrow -\infty ,  \label{27}
\end{equation}

Now we can proceed to solve Eq. (20) in the near horizon limit and then
impose the above boundary condition to find the frequency of QNM i.e., $%
\omega $. Expansion of the metric function $f(r)$ around the event horizon
is given by

\begin{align}
f(r)& =f^{\prime }(r_{h})(r-r_{h})+\Game (r-r_{h})^{2},  \notag \\
& \simeq 2\kappa (r-r_{h}),  \label{28}
\end{align}

where $\kappa $\ is the surface gravity, which is nothing but $\frac{1}{2}%
f^{\prime }(r_{h})$. From Eq. (22) we now obtain

\begin{equation}
r^{\ast }\simeq \frac{1}{2\kappa }\ln (r-r_{h}),  \label{29}
\end{equation}

Furthermore, after letting $x=r-r_{+}$ and inserting Eq. (28) into Eq. (21)
together with performing Taylor expansion around $x=0$, one gets the near
horizon form of the effective potential as,

\begin{equation}
V(x)\simeq 2\kappa x\left[ \frac{L(L+1)}{\gamma ^{2}r_{h}^{2\acute{N}}}(1-%
\frac{2\acute{N}x}{r_{h}})+\frac{2\acute{N}\kappa }{r_{h}}(1-\frac{x}{r_{h}}%
)+\frac{2\acute{N}\kappa x}{r_{h}^{2}}(\acute{N}-1)\right] ,  \label{30}
\end{equation}

After substituting Eq. (30) into the Zerilli equation (20), we find

\begin{equation}
-4\kappa ^{2}x^{2}\frac{d^{2}\rho (x)}{dx^{2}}-4\kappa ^{2}x\frac{d\rho (x)}{%
dx}+V(x)\rho (x)=\omega ^{2}\rho (x),  \label{31}
\end{equation}

Solution of the above equation admits

\begin{equation}
\rho (x)\sim \varepsilon ^{\frac{i\omega }{2\kappa }}U(a,b,c),  \label{32}
\end{equation}

where $U(a,b,c)$\ is the confluent hypergeometric function \cite{Abramowitz}%
. The parameters of the confluent hypergeometric functions are found as

\begin{align}
a& =\frac{1}{2}+i(\frac{\omega }{2\kappa }-\frac{\hat{\alpha}}{\hat{\beta}%
\sqrt{\kappa }\gamma }),  \notag \\
b& =1+i\frac{\omega }{\kappa },  \label{33} \\
c& =i\frac{\hat{\beta}x}{2\gamma r_{h}\sqrt{\kappa }},  \notag
\end{align}

where

\begin{align}
\hat{\beta}& =4r_{h}^{(\acute{N}-\frac{1}{2})}\sqrt{\acute{N}L(L+1)+\acute{N}%
\kappa \gamma ^{2}(2-\acute{N})r_{h}^{(2\acute{N}-1)}},  \notag \\
\hat{\alpha}& =L(L+1)+2\acute{N}\kappa \gamma ^{2}r_{h}^{(2\acute{N}-1)},
\label{34}
\end{align}

One can easily check that these results are in consistent with the studies
done for the $4D$ LDBH ($\acute{N}=\frac{1}{2}$) \cite{Sakalli2}.

In the limit of $x\ll 1$, the solution (32) becomes

\begin{equation}
\rho (x)\sim c_{1}x^{-\frac{i\omega }{2\kappa }}\frac{\Gamma (i\frac{\omega 
}{\kappa })}{\Gamma (a)}+c_{2}x^{\frac{i\omega }{2\kappa }}\frac{\Gamma (-i%
\frac{\omega }{\kappa })}{\Gamma (1+a-b)},  \label{35}
\end{equation}

where constants $c_{1}$ and $c_{2}$ denote the amplitudes of the
near-horizon outgoing and ingoing waves, respectively. Now, since there is
no outgoing wave in the QNM at the horizon, the first term of Eq. (35)
should be vanished. This is possible with the poles of the Gamma function of
the denominator. Therefore, the poles of the Gamma function are the decision
makers of the frequencies of the QNMs. Thus, we can read the frequencies of
the QNMs of the CDBHs as,

\begin{equation}
\omega _{\tilde{n}}=\frac{2\sqrt{\kappa }\hat{\alpha}}{\hat{\beta}\gamma }%
+i(2\tilde{n}+1)\kappa ,\text{ \ \ \ \ \ \ }(\tilde{n}=1,2,3,...)  \label{36}
\end{equation}

where $\tilde{n}$\ is the overtone quantum number of the QNM. Thus, the
imaginary part of the frequency of the QNM is

\begin{equation}
\omega _{I}=(2\tilde{n}+1)\kappa =\frac{2\pi }{\hbar }(2\tilde{n}+1)T_{H},
\label{37}
\end{equation}

where $T_{H}=\frac{\hbar \kappa }{2\pi }$ which is called the Hawking
temperature \cite{Rovelli}. Hence the transition frequency between two
highly damped neighboring states becomes $\Delta \omega =\omega _{\tilde{n}%
+1}-\omega _{\tilde{n}}=4\pi T_{H}$. So the adiabatic invariant quantity (2)
in this case results with

\begin{equation}
I_{adb}=\frac{\hbar }{4\pi }\int \frac{dM}{T_{H}},  \label{38}
\end{equation}

Recalling the first law of thermodynamics (16), we easily see that

\begin{equation}
I_{adb}=\frac{S_{BH}}{4\pi }\hbar ,  \label{39}
\end{equation}

Finally, according to the Bohr-Sommerfeld quantization rule $I_{adb}=\hbar
n, $ one gets the spacing of the entropy spectrum as

\begin{equation}
S_{n}=4\pi n,  \label{40}
\end{equation}

Since $S=\frac{\mathcal{A}}{4\hbar },$ the area spectrum is obtained as

\begin{equation}
\mathcal{A}_{n}=16\pi n\hbar ,  \label{41}
\end{equation}

From the above, we can simply measure the area spacing as

\begin{equation}
\Delta \mathcal{A}=16\pi \hbar .  \label{42}
\end{equation}

It is easily seen that the spectroscopy of the CDBH is completely
independent of the dilaton parameter $a$. Besides, the spacings between the
levels are double of the Bekenstein's original result. This means that $%
\epsilon =16\pi $. The discussion on this differentness is made in the
conclusion part. Nevertheless, the obtained entropy and area spectra are
evenly spaced. The latter result is in agreement with the Wei et al.'s
conjecture \cite{Samp3} which proposes that static BHs of Einstein's gravity
theory has equidistant quantum spectra of both entropy and area .

\section{Conclusion}

In this paper, the quantum spectra of the CDBH are investigated through the
MM, which is based on adiabatic invariance of BHs. In order to obtain the
QNM of the CDBH, we applied an approximation method given in Refs. \cite%
{Appro1,Appro2,Appro3} to the Zerilli equation (20). After a straightforward
calculation, by using the MM which employs the proper frequency as the
imaginary part instead of the real part of the QNMs the entropy and area
spectra of the CDBH are derived. Both spectra are independent of the dilaton
parameter and equally spaced as such as in the case of the LDBH \cite%
{Sakalli2}. However, we obtained $\epsilon =16\pi $ which results that the
equi-spacing is different than its usual Schwarzschild value: $\epsilon
=8\pi $. This discrepancy may arise due to the Schwinger mechanism \cite%
{SKim}. Because, in the Bekenstein's original work \cite{Bek2}, one gets the
entropy spectrum by combining both the Schwinger mechanism and the
Heisenberg quantum uncertainty principle. However, the QNM method that
applied herein considers only the uncertainty principle via the
Bohr-Sommerfeld quantization (40). Therefore, as stated in Ref. \cite{Hod2},
the spacings between two neighboring levels may become different depending
on the which method is applied. Thus, getting $\epsilon =16\pi $ rather than
its usual value $\epsilon =8\pi $ is not suprising. Finally, we would like
to point out that it will be interesting to apply the same analysis to the
other dilatonic BHs like the dyonic BHs \cite{Yazad,Yazad2}. This is going
to be our next problem in the near future.

\end{document}